\documentclass[twocolumn,showpacs,aps,prl,superscriptaddress,reprint]{revtex4}

\usepackage{graphicx}
\usepackage{dcolumn}
\usepackage{bm}

\begin{document}

\title{Measurement of the parity-violating asymmetry in inclusive
  electroproduction of $\pi^-$ near the $\Delta^0$ resonance}

\author{D.~Androi\'c}
\affiliation{Department of Physics, University of Zagreb, Zagreb HR-41001 Croatia} 

\author{D.~S.~Armstrong}
\affiliation{Department of Physics, College of William and Mary, Williamsburg, VA 23187 USA} 

\author{J.~Arvieux$^\dagger$}
\affiliation{Institut de Physique Nucl\'eaire d'Orsay, Universit\'e Paris-Sud, F-91406 Orsay Cedex FRANCE}

\author{S.~L.~Bailey}
\affiliation{Department of Physics, College of William and Mary, Williamsburg, VA 23187 USA} 

\author{D.~H.~Beck}
\affiliation{Loomis Laboratory of Physics, University of Illinois, Urbana, IL 61801 USA}

\author{E.~J.~Beise}
\affiliation{Department of Physics, University of Maryland, College Park, MD 20742 USA}

\author{J.~Benesch}
\affiliation{Thomas Jefferson National Accelerator Facility, Newport News, VA 23606 USA}

\author{F.~Benmokhtar}
\affiliation{Department of Physics, University of Maryland, College Park, MD 20742 USA}
\affiliation {Department of Physics, Carnegie Mellon University, Pittsburgh, PA 15213 USA}

\author{L.~Bimbot}
\affiliation{Institut de Physique Nucl\'eaire d'Orsay, Universit\'e Paris-Sud, F-91406 Orsay Cedex FRANCE}

\author{J.~Birchall}
\affiliation{Department of Physics, University of Manitoba, Winnipeg, MB R3T 2N2 CANADA}

\author{P.~Bosted}
\affiliation{Thomas Jefferson National Accelerator Facility, Newport News, VA 23606 USA}

\author{H.~Breuer}
\affiliation{Department of Physics, University of Maryland, College Park, MD 20742 USA}

\author{C.~L.~Capuano}
\affiliation{Department of Physics, College of William and Mary, Williamsburg, VA 23187 USA} 

\author{Y.-C.~Chao}
\affiliation{Thomas Jefferson National Accelerator Facility, Newport News, VA 23606 USA}

\author{A.~Coppens}
\affiliation{Department of Physics, University of Manitoba, Winnipeg, MB R3T 2N2 CANADA} 

\author{C.~A.~Davis}
\affiliation{TRIUMF, Vancouver, BC V6T 2A3 CANADA}

\author{C.~Ellis}
\affiliation{Department of Physics, University of Maryland, College Park, MD 20742 USA}

\author{G.~Flores}
\affiliation{Department of Physics, New Mexico State University, Las Cruces, NM 88003 USA}

\author{G.~Franklin}
\affiliation {Department of Physics, Carnegie Mellon University, Pittsburgh, PA 15213 USA}

\author{C.~Furget}
\affiliation{LPSC, Universit\'e Joseph Fourier Grenoble 1, CNRS/IN2P3, Institut  Polytechnique de Grenoble, Grenoble, FRANCE}

\author{D.~Gaskell}
\affiliation{Thomas Jefferson National Accelerator Facility, Newport News, VA 23606 USA}

\author{M.~T.~W.~Gericke}
\affiliation{Department of Physics, University of Manitoba, Winnipeg, MB R3T 2N2 CANADA}

\author{J.~Grames}
\affiliation{Thomas Jefferson National Accelerator Facility, Newport News, VA 23606 USA}

\author{G.~Guillard}
\affiliation{LPSC, Universit\'e Joseph Fourier Grenoble 1, CNRS/IN2P3, Institut  Polytechnique de Grenoble, Grenoble, FRANCE}

\author{J.~Hansknecht}
\affiliation{Thomas Jefferson National Accelerator Facility, Newport News, VA 23606 USA}

\author{T.~Horn}
\affiliation{Thomas Jefferson National Accelerator Facility, Newport News, VA 23606 USA}

\author{M.~K.~Jones}
\affiliation{Thomas Jefferson National Accelerator Facility, Newport News, VA 23606 USA}

\author{P.~M.~King}
\affiliation{Department of Physics and Astronomy, Ohio University, Athens, OH 45701 USA}

\author{W.~Korsch}
\affiliation{Department of Physics and Astronomy, University of Kentucky, Lexington, KY 40506 USA}

\author{S.~Kox}
\affiliation{LPSC, Universit\'e Joseph Fourier Grenoble 1, CNRS/IN2P3, Institut  Polytechnique de Grenoble, Grenoble, FRANCE}

\author{L.~Lee}
\affiliation{Department of Physics, University of Manitoba, Winnipeg, MB R3T 2N2 CANADA}

\author{J.~Liu}
\affiliation{Kellogg Radiation Laboratory, California Institute of Technology,  Pasadena, CA 91125 USA}

\author{A.~Lung}
\affiliation{Thomas Jefferson National Accelerator Facility, Newport News, VA 23606 USA}

\author{J.~Mammei}
\affiliation{Department of Physics, Virginia Tech, Blacksburg, VA 24061 USA}

\author{J.~W.~Martin}
\affiliation{Department of Physics, University of Winnipeg, Winnipeg, MB R3B 2E9 CANADA}

\author{R.~D.~McKeown}
\affiliation{Kellogg Radiation Laboratory, California Institute of Technology,  Pasadena, CA 91125 USA}

\author{A.~Micherdzinska}
\affiliation{Department of Physics, The George Washington University, Washington, DC 20052 USA}

\author{M.~Mihovilovic}
\affiliation{Jo\^zef Stefan Institute, 1000 Ljubljana, SLOVENIA}

\author{H.~Mkrtchyan}
\affiliation{Yerevan Physics Institute, Yerevan 375036 ARMENIA}

\author{M.~Muether}
\affiliation{Loomis Laboratory of Physics, University of Illinois, Urbana, IL 61801 USA}

\author{S.~A.~Page}
\affiliation{Department of Physics, University of Manitoba, Winnipeg, MB R3T 2N2 CANADA}

\author{V.~Papavassiliou}
\affiliation{Department of Physics, New Mexico State University, Las Cruces, NM 88003 USA}

\author{S.~F.~Pate}
\affiliation{Department of Physics, New Mexico State University, Las Cruces, NM 88003 USA}

\author{S.~K.~Phillips}
\affiliation{Department of Physics, College of William and Mary, Williamsburg, VA 23187 USA} 

\author{P. Pillot}
\affiliation{LPSC, Universit\'e Joseph Fourier Grenoble 1, CNRS/IN2P3, Institut  Polytechnique de Grenoble, Grenoble, FRANCE}

\author{M.~L.~Pitt}
\affiliation{Department of Physics, Virginia Tech, Blacksburg, VA 24061 USA}

\author{M.~Poelker}
\affiliation{Thomas Jefferson National Accelerator Facility, Newport News, VA 23606 USA}

\author{B.~Quinn}
\affiliation {Department of Physics, Carnegie Mellon University, Pittsburgh, PA 15213 USA}

\author{W.~D.~Ramsay}
\affiliation{Department of Physics, University of Manitoba, Winnipeg, MB R3T 2N2 CANADA}
\affiliation{TRIUMF, Vancouver, BC V6T 2A3 CANADA}

\author{J.-S.~Real}
\affiliation{LPSC, Universit\'e Joseph Fourier Grenoble 1, CNRS/IN2P3, Institut  Polytechnique de Grenoble, Grenoble, FRANCE}

\author{J.~Roche}
\affiliation{Department of Physics and Astronomy, Ohio University, Athens, OH 45701 USA}

\author{P.~Roos}
\affiliation{Department of Physics, University of Maryland, College Park, MD 20742 USA}

\author{J.~Schaub}
\affiliation{Department of Physics, New Mexico State University, Las Cruces, NM 88003 USA}

\author{T.~Seva}
\affiliation{Department of Physics, University of Zagreb, Zagreb HR-41001 Croatia} 

\author{N.~Simicevic}
\affiliation{Department of Physics, Louisiana Tech University,  Ruston, LA 71272 USA}

\author{G.~R.~Smith}
\affiliation{Thomas Jefferson National Accelerator Facility, Newport News, VA 23606 USA}

\author{D.~T.~Spayde}
\affiliation{Department of Physics, Hendrix College, Conway, AR 72032 USA}

\author{M.~Stutzman}
\affiliation{Thomas Jefferson National Accelerator Facility, Newport News, VA 23606 USA}

\author{R.~Suleiman}
\affiliation{Department of Physics, Virginia Tech, Blacksburg, VA 24061 USA}
\affiliation{Thomas Jefferson National Accelerator Facility, Newport News, VA 23606 USA}

\author{V.~Tadevosyan}
\affiliation{Yerevan Physics Institute, Yerevan 375036 ARMENIA}

\author{W.~T.~H.~van~Oers}
\affiliation{Department of Physics, University of Manitoba, Winnipeg, MB R3T 2N2 CANADA}

\author{M.~Versteegen}
\affiliation{LPSC, Universit\'e Joseph Fourier Grenoble 1, CNRS/IN2P3, Institut  Polytechnique de Grenoble, Grenoble, FRANCE}

\author{E.~Voutier}
\affiliation{LPSC, Universit\'e Joseph Fourier Grenoble 1, CNRS/IN2P3, Institut  Polytechnique de Grenoble, Grenoble, FRANCE}

\author{W.~Vulcan}
\affiliation{Thomas Jefferson National Accelerator Facility, Newport News, VA 23606 USA}

\author{S.~P.~Wells}
\affiliation{Department of Physics, Louisiana Tech University,  Ruston, LA 71272 USA}

\author{S.~E.~Williamson}
\affiliation{Loomis Laboratory of Physics, University of Illinois, Urbana, IL 61801 USA}

\author{S.~A.~Wood}
\affiliation{Thomas Jefferson National Accelerator Facility, Newport News, VA 23606 USA}

\collaboration{G0 Collaboration}

\date{\today}

\begin{abstract}
The parity-violating (PV) asymmetry of inclusive $\pi^-$ production in
electron scattering from a liquid deuterium target was measured at
backward angles.  The measurement was conducted as a part of the G0
experiment, at a beam energy of 360~MeV.  The physics process
dominating pion production for these kinematics is quasi-free
photoproduction off the neutron via the $\Delta^0$ resonance.  In the
context of heavy-baryon chiral perturbation theory (HB$\chi$PT), this
asymmetry is related to a low energy constant $d_\Delta^-$ that
characterizes the parity-violating $\gamma$N$\Delta$ coupling.  Zhu et
al.~calculated $d_\Delta^-$ in a model benchmarked by the large
asymmetries seen in hyperon weak radiative decays, and predicted
potentially large asymmetries for this process, ranging from
$A_\gamma^-=-5.2$ to $+5.2$~ppm.  The measurement performed in this
work leads to $A_\gamma^-=-0.36\pm 1.06\pm 0.37\pm 0.03$~ppm (where
sources of statistical, systematic and theoretical uncertainties are
included), which would disfavor enchancements considered by Zhu et
al.~proportional to $V_{ud}/V_{us}$.  The measurement is part of a
program of inelastic scattering measurements that were conducted by
the G0 experiment, seeking to determine the $N-\Delta$ axial
transition form-factors using PV electron scattering.
\end{abstract}


\maketitle
 
In electron scattering, the size of parity-violating asymmetries is
usually related to an interference between Z and $\gamma$ exchange
amplitudes.  Therefore, in the photoproduction limit ($Q^2=0$, where
$Q^2$ is the negative four-momentum transfer squared) virtual Z bosons
cannot be exchanged, and the asymmetry is expected to tend to zero.
But, in the case of scattering from nucleons, parity-violation also
occurs in weak interactions among quarks, generically referred to as
the hadronic weak interaction; this form of an electroweak radiative
correction can lead to non-zero asymmetries in the photoproduction
limit.

Zhu {\it et al.}~\cite{Zhu1,Zhu2} studied electroweak radiative
corrections in the photoproduction limit theoretically for PV
inelastic scattering of electrons from nucleons.  The variation of the
PV asymmetry with $Q^2$ in this case is particularly of interest
because of the desire to extract $N-\Delta$ axial transition form
factors, and to compare to and improve the determinations made by
neutrino scattering experiments.

In Ref.~\cite{Zhu1}, the PV asymmetry $A_\gamma^-$ was calculated for
the process $\vec{\gamma}+d\rightarrow\Delta^0+p\rightarrow\pi^-+p+p$
using HB$\chi$PT.  The PV asymmetry was found to be related to a new
low-energy constant in the effective weak Lagrangian $d_\Delta^-$
characterizing the PV $\gamma N \Delta$ coupling:
\begin{equation}
A_{\gamma}^-\equiv\frac{d\sigma_R-d\sigma_L}{d\sigma_R+d\sigma_L}=-\frac{2d_\Delta^{-}}{C_3^V}\frac{M_N}{\Lambda_\chi}
\label{eqn:zhu}
\end{equation} 
where $d\sigma_{R,L}$ are the differential cross sections for right-
(R) or left- (L) circular-polarized incident photons, $C_3^V$ is the
dominant $N-\Delta$ vector transition form factor, $M_N$ is the mass
of the nucleon, and $\Lambda_\chi$ is the scale of chiral symmetry
breaking.  Non-resonant, higher order chiral, and $1/M_N$ corrections
are neglected here, as in Ref.~\cite{Zhu1}.

By naive dimensional analysis, it would be expected that
$d_\Delta^\pm\sim g_\pi$, where $g_\pi\sim G_FF_\pi^2/2\sqrt{2}\sim
5\times10^{-8}$ is the scale of the weak charged-current hadronic
process.  (Here, the quantity $d_\Delta^+$ refers specifically to the
process $\vec{\gamma}+p\rightarrow\Delta^+\rightarrow\pi^++n$.) Zhu et
al.~considered possible enhancements to $d_\Delta^\pm$ via the
inclusion of intermediate $J^\pi=\frac{1}{2}^-$ and $\frac{3}{2}^-$
resonances which would mix with the nucleon or $\Delta$ respectively
via the hadronic weak interaction.  A similar treatment yielded
excellent agreement with observables in hyperon decay, simultaneously
describing weak radiative and weak hadronic decay in the $\Delta S=1$
sector of the hadronic weak interaction \cite{zhu_ref_24_25}.  In the
$\Delta S=0$ sector, less information about the amplitudes was known,
and so their scale was taken to be of order the $\Delta S=1$
amplitudes.  Due to unknown possible phase factors between the
amplitudes, this resulted in a range of predictions of
$|d_\Delta^\pm|\sim(10-25)g_\pi$.  Since the amplitudes could be
related to the hadronic charged-current interaction, the $\Delta S=0$
amplitudes might be further enhanced over their $\Delta S=1$
counterparts by a factor $V_{ud}/V_{us}$ where $V_{ij}$ are elements
of the CKM matrix .

Zhu {\it et al.}~concluded that a reasonable range of predictions is
$|d_\Delta^\pm|=(1-100)g_\pi$, citing a ``best value'' of
$|d_\Delta^\pm|=25g_\pi$ \cite{Zhu1}.  Through equation
(\ref{eqn:zhu}), this corresponds to a range
$|A_\gamma^\pm|=(0.052-5.2)$~ppm, with a best value of 1.3~ppm.  We
sought to test these predictions experimentally via inclusive
electroproduction of $\pi^-$ off the deuteron.  The detailed analysis
is presented in Ref.~\cite{bib:coppens}.

Data were acquired during the backward angle phase of the G0
experiment, performed in Hall C at Jefferson Laboratory
\cite{G0BackPaper}.  Data were acquired during our low-energy liquid
deuterium (LD$_2$) target measurements, collected simultaneously with
inclusive quasi-elastic and inelastic electron scattering data, over a
two-week period.

The G0 experimental apparatus was described in Ref.~\cite{bib:g0exp}.
A polarized electron beam of current 35~$\mu$A and energy 360~MeV
impinged on a 20~cm liquid deuterium target~\cite{G0targ}.  The
average beam polarization, measured with M\o ller~\cite{bib:moller}
and Mott polarimeters \cite{spindance}, was $85.8 \pm 2.1$\% (combined
statistical and systematic uncertainty).  Helicity-correlated current
changes were corrected with an active feedback system.

A superconducting toroidal spectrometer, consisting of an eight-coil
magnet, collimators, and eight detector sets, detected $\pi^-$
scattered at an average angle of 100$^\circ$.  Each detector set
included two arrays of scintillators, one near the exit of the magnet
(``CED''), and the second along its focal surface (``FPD'').  For each
detector set, an aerogel \v{C}erenkov detector with a pion threshold
of 570 MeV was used in concert with the scintillators, allowing
separation of $\pi^-$ from electrons.

\begin{figure}
\resizebox{20.5pc}{!}{\includegraphics{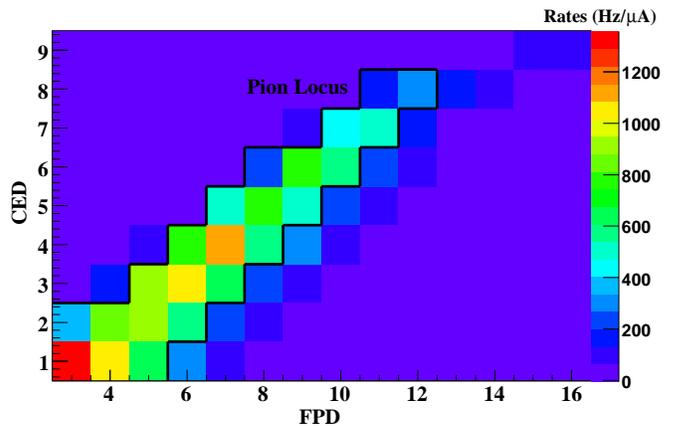}}
\caption{\label{fig:matrix} (color online).  Pion counting rates for
  various CED-FPD combinations (FPDs 1 and 2 not used).  Pion tracks
  occur mainly in the lower left corner of the matrix.  Rates in the
  far upper right corner are due to misidentification of electrons via
  \v{C}erenkov ineffiency for electron detection.  The locus of
  combinations analyzed for inclusive pion production is outlined in
  black.}
\end{figure}

The pion rate was signified by a coincidence between particular pairs
of CED's and FPD's, and an absence of a signal above threshold in the
\v{C}herenkov detector.  Rates were corrected for deadtime and random
coincidences in a manner analogous to our electron
data~\cite{G0BackPaper}.  The overall deadtime for this dataset was
15\%.  The rate sensed within the selected locus of CED-FPD pairs is
displayed graphically in Fig.~\ref{fig:matrix}.

Helicity-correlated beam properties were characterized using
beam-current and beam-position monitors.  Sensitivities of the
detector to changes in beam current, position, angle, and energy were
also measured.  Instead of correcting for helicity-correlated beam
properties, a conservative error of 0.21~ppm was assigned, determined
by multiplying the largest observed helicity correlated property times
the largest sensitivity, averaged separately over the run period.
Averaging the product of the two appropriately would have resulted in
negligible overall corrections.

Possible electronic leakage of the helicity signal into the
data-acquisition system was studied by periodically inserting a
half-wave plate into the laser beam path in the polarized source,
which would act to reverse the direction of the polarization of
incident electrons.  Upon insertion, all asymmetries measured by the
experiment should reverse sign, and averaging the results for
different half-wave plate states should result in zero.  In the case
of these data, the average determined in this way showed some lack of
consistency across octants.  Averaging in turn over octants gave
$(1.7\pm 0.8)$~ppm (prior to correction for beam polarization), in
reasonable agreement with zero.  No evidence of an unknown systematic
effect could be found subdividing the data in different ways and in
particular studying known octant-dependent corrections.
Furthermore, the data, when the correct half-wave plate setting was
taken into account, were statistically consistent.  Therefore no
additional systematic uncertainty was assigned.

\begin{figure}
\resizebox{20.5pc}{!}{\includegraphics{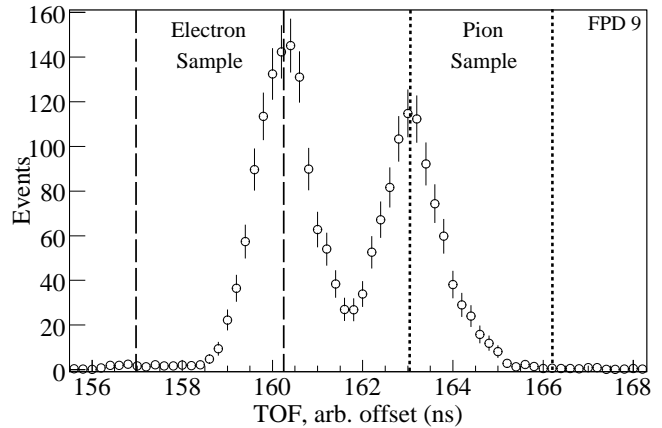}}
\caption{\label{fig:3.12} (color online).  Sample time-of-flight
  spectrum (for particles sensed by detector FPD09) during a
  data-taking run using a pulsed beam.  Electrons correspond to the
  left peak and pions to the right peak.  Analysis of the
  time-spectrum in concert with the \v{C}erenkov counter enabled a
  measurement of the particle identification properties of the
  \v{C}erenkov counter.}
\end{figure}

The data were then corrected for backgrounds.  In these data,
backgrounds were mainly due to misidentified electrons which did not
create a signal above threshold in the \v{C}erenkov detector.  The
backgrounds were characterized in special data-taking runs where the
electron beam was pulsed at 31 MHz.  In these runs, time-of-flight
(TOF) spectra for particles (their flight path being from the target
to the FPD's) were used as an alternate method to determine the
particles' identities.  By defining hard cuts on TOF, pure samples of
pions and electrons could be defined, which would then be used to
characterize \v{C}erenkov performance (see Fig.~\ref{fig:3.12}).
Particle fluxes could be estimated from two-Gaussian fits to the TOF
spectrum.  The combination of techniques allowed determination of the
pion efficiency and electron contamination for the pion sample.

The CED-FPD pairs in the pion locus were selected by requiring the
electron contamination of the pion sample, in a given pair, before
background correction, to be below 10\%.  The resultant average
contamination by electrons for the pion locus was 2.6\%.  This was
corrected by appropriately subtracting the measured electron asymmetry
in each of the same CED-FPD pairs.  (Without the veto provided by the
\v{C}erenkov counter, the electron contamination would have been
20\%.)  The average efficiency for pion identification was $>99\%$ for
pions satisfying the CED-FPD coincidence condition.


The polarization axis of the electron beam was controlled by a Wien
filter in the 5 MeV section of the accelerator.  The Wien filter
setting was adjusted to optimize the longitudinal polarization in
dedicated measurements with the M\o ller polarimeter in Hall C
\cite{bib:moller}.  The resultant beam, while dominantly polarized in
the longitudinal direction, possessed a slight degree of polarization
transverse to the direction of propagation, in the bend plane of the
accelerator.  This in turn resulted in a parity-conserving azimuthal
dependence to the asymmetries measured by the experiment, which can be
sensed because of the azimuthal segmentation of the detectors into
octants.

By adjusting the Wien filter setting, dedicated runs were conducted
with the degree of transverse polarization arranged to be as large as
possible, so that the sensitivity of the detector to this azimuthal
asymmetry could be deduced.  The azimuthal asymmetry measured by this
technique was sinusoidal in its dependence over octants, with an
amplitude of $\sim 170$~ppm.  It is believed that this rather strong
azimuthal dependence ultimately results from a sensitivity to the LT'
interference term seen in parity-conserving pion electroproduction
\cite{bib:harryprivate}, and we intend to study this process in a
separate publication.

\begin{figure}
\resizebox{20.5pc}{!}{\includegraphics{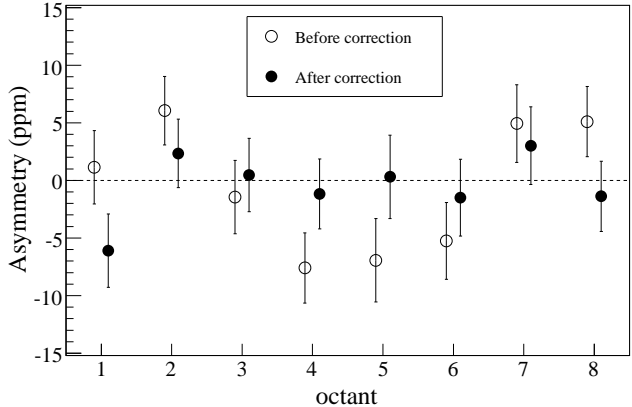}}
\caption{\label{fig:3.11+3.18} Measured pion asymmetries before and
  after correction for the residual transverse polarization in the
  electron beam.  The correction successfully diminishes the
  sinusoidal variation with octant number from the data, with little
  impact on the octant-averaged result for the asymmetry, and reduces
  $\chi^2$/DOF for a fit to a flat line from 3.5 to 1.2.}
\end{figure}

The luminosity monitors for the experiment \cite{bib:g0exp} were
also segmented azimuthally.  By comparing the luminosity monitor
asymmetry under the transverse and longitudinal Wien filter settings,
the degree of transverse polarization in the nominally longitudinal
beam was deduced to be $4.3\pm 0.2$\%.  Using the azimuthal pion
asymmetries determined for transversely polarized beam, and the degree
of transverse polarization measured using the luminosity monitors, the
pion longitudinal asymmetries could be corrected as a function of
octant.  The success of this correction in removing the residual
azimuthal dependence in the nominally longitudinally polarized
electron beam data is displayed graphically in
Fig.~\ref{fig:3.11+3.18}.

Backgrounds due to the thin aluminum target windows were 2\%.  These
were not corrected because quasi-free production of $\pi^-$ off
neutrons is expected to dominate the asymmetry, and therefore this
process should carry the same asymmetry as the deuterium data to well
within the precision of the data.  Correcting finally for beam
polarization results in the measured raw PV asymmetry attributable to
inclusive production of $\pi^-$ off the LD$_2$ target.  The inclusive
pion asymmetry including all experimental corrections was $A_{\rm
  meas}=-0.55\pm 1.03\pm 0.37$~ppm, where the first uncertainty is
statistical, and the second systematic.  A list of the systematic
uncertainties is presented in Table~\ref{tab:systematics}.



\begin{table}
\caption{\label{tab:systematics} Systematic uncertainties assigned for
  the corrections described in the text.}
\begin{ruledtabular}
\begin{tabular}{lc}
Source & Uncertainty (ppm)\\
\hline
Rate corrections & 0.26  \\
Helicity-correlated corrections & 0.21 \\
Backgrounds, particle identification & 0.12\\
Residual transverse polarization & 0.08 \\
Beam Polarization & 0.01 \\
\end{tabular}
\end{ruledtabular}
\end{table}
 
The measured asymmetry, $A_{\rm meas}$, includes pion fluxes induced
by both photoproduction and electroproduction of pions.  We desire to
extract the photoproduction asymmetry $A_{\gamma}^-$ that would be
induced by incident real photons.  The two are related by:
\begin{equation}
A_{\rm meas}=f_{\rm brems}\langle D(y)\rangle A_\gamma^-+f_{\rm
  virt}\langle A_e^-(Q^2)\rangle .
\label{eqn:agamma}
\end{equation}
Here, $f_{\rm brems}$ and $f_{\rm virt}$ are the fractional fluxes of
pions initiated by bremsstrahlung photons and virtual photons
(i.e. electroproduction), respectively ($f_{\rm brems}+f_{\rm
  virt}=1$).  The factor $D(y)$ is the degree of circular polarization
carried by the bremsstrahlung beam \cite{bib:alex125}, relative to the
electron beam ($y$ being the fractional energy carried by the photon).
The factor $A_e^-(Q^2)$ is the asymmetry for electroproduction of
pions.  According to theoretical expectation \cite{Zhu2}, $A_e^-(Q^2)$
is approximately linear in $Q^2$ for the range of $Q^2$ dominating
this experiment, with the intercept at $Q^2=0$ being equal to
$A_\gamma^-$.  We therefore characterized the average $Q^2$ for the
electroproduction events and extrapolate to the photon point.  Since
the scattered electrons were not detected in coincidence with the
scattered pions, for this measurement, we employed simulation
techniques to calculate the factors $f_{\rm virt}$, $\langle
D(y)\rangle$, and $\langle Q^2\rangle$.  Theoretical input was used to
constrain the slope of the electroproduction asymmetry with $Q^2$.

The simulation was benchmarked by comparing with our measured pion
rates, and their distribution in the acceptance of the experiment.
The simulation of the detector acceptance was based on the GEANT3
toolkit \cite{bib:geant3}.  Pion absorption in the apparatus was
estimated to affect the rate at the percent level, and would not
affect the PV asymmetry determination.  Physics generators for both
bremsstrahlung and virtual photon induced reactions were developed.
The cross-sections used in the generators were based on the MAID model
of pion production \cite{bib:maid}, applied to a neutron target.
These were further tested by comparing with published extractions of
the $n(\gamma,\pi^-)$ process, which were based on measurements of
$d(\gamma,\pi^-)pp_s$ \cite{bib:benz}, and found to be in good
agreement.  Nuclear corrections to the cross section were based on the
same reference.  Additionally, corrections for Fermi motion were
computed by generating a random initial-state neutron momentum
according to a parametrization of the nucleon momentum distribution in
the deuteron, and were found to smear the pion rates in the detector
acceptance.  (Possible nuclear corrections to the PV asymmetry were
argued to be small in Ref.~\cite{Zhu1}, and therefore we made no
correction for such effects.)  Particular care was taken in the
generation of electroproduction events, where virtual-photon flux
formulae valid down to $Q^2\sim m_e^2$ were used.  This part of the
cross-section was also compared with analytical formulae
\cite{bib:tiatorwright} over a broad range of kinematics.

The simulation of the pion rate was found to agree with the data to
within 15\%, generally reproducing trends seen in CED-FPD space.  The
fraction of events induced by virtual photons was found to be $f_{\rm
  virt}=0.45\pm 0.07$, in good agreement with simple estimates based
on the target's radiation length and the effective radiation length
for virtual photon induced reactions.  The uncertainty was assigned
based on the level of agreement of the simulation with data, and with
the simple estimates.

The average $\langle D(y)\rangle$ was $0.95\pm 0.05$, where the stated
uncertainty is systematic.  The quantity $D(y)$ becomes unity as
$y\rightarrow 1$ and for 90\% of the simulated events, $y>0.7$
corresponding to $D(y)>0.9$~\cite{bib:alex125}.  We therefore think
the assigned systematic uncertainty is conservative.  The average
accepted photon energy was 320~MeV ($y=0.89$); the average invariant
mass of the final-state hadronic system was $W=$1220~MeV.

The average $\langle Q^2\rangle$ for electroproduction events was
determined to be $0.0032$~(GeV/$c$)$^2$.  A systematic uncertainty of
10\% on $\langle Q^2\rangle$ was assigned based on shifts observed in
the simulation varying the magnetic field, beam energy, and target
position within reasonable ranges.  This agreed to the same level of
precision with a simple estimate based on the virtual photon flux
factor varying approximately as $1/Q^2$ and averaging over the
permitted electron kinematics.  The simulated electroproduction events
were heavily weighted towards low $Q^2$ with 90\% of them falling
below $Q^2=0.01$~(GeV/$c$)$^2$.

The slope of $A_e^-(Q^2)$ with $Q^2$ was estimated based on
Ref.~\cite{Zhu2}.  The dominant term in the slope is a constant
related linearly to $\sin^2\theta_W$.  The slope was assigned a 14\%
theoretical uncertainty, which is the full size of the non-resonant
and structure-dependent terms in the asymmetry, including the
electroweak radiative corrections, calculated in the same reference.

Solving equation~(\ref{eqn:agamma}) then yields $A_\gamma^-=-0.36\pm
1.06\pm 0.37\pm 0.03$~ppm where the third uncertainty is the theory
uncertainty explained above.  Using equation~(\ref{eqn:zhu}) with the
values $C_3^V=1.6$ and $\Lambda_\chi=1$~GeV (from Ref.~\cite{Zhu2})
then yields $d_\Delta^-=(8.1\pm 23.7\pm 8.3\pm 0.7)g_\pi$ where
$g_\pi=3.8\times 10^{-8}$.  No additional uncertainty was assigned
for the interpretation in this particular model.

Our new result means that possible enhancements considered in
Ref.~\cite{Zhu1}, proportional to $V_{ud}/V_{us}$, are disfavored.
The possibility of an unexpectedly large PV asymmetry in pion
photoproduction on the $\Delta$-resonance has been limited to the ppm
level.

Results on related parameters in PV inclusive inelastic electron
scattering are forthcoming from the G0 experiment \cite{bib:neven} and
will be related in a separate publication \cite{bib:carissa}.
Measurements being conducted by the $Q_{weak}$ experiment
\cite{bib:qweak} will shed light on the inclusive parameter $d_\Delta$
via PV inclusive inelastic electron scattering at low $Q^2\sim
0.027$~(GeV/$c$)$^2$.


\begin{acknowledgments}
We gratefully acknowledge the strong technical contributions to this
experiment from many groups: Caltech, Illinois, LPSC-Grenoble,
IPN-Orsay, TRIUMF and particularly the Accelerator and Hall C groups
at Jefferson Lab.  CNRS (France), DOE (U.S.), NSERC (Canada) and NSF
(U.S.) supported this work in part.
\end{acknowledgments}

\end{document}